# E-BUSINESS IMPLICATIONS FOR PRODUCTIVITY AND COMPETITIVENESS

**Pece Mitrevski, PhD in Computer Science**

**Faculty of Technical Sciences, Bitola, Macedonia**

**Olivera Kostoska, MSc in Economics**

**Faculty of Economics, Prilep, Macedonia**

**Marjan Angeleski, MBA**

**Faculty of Economics, Prilep, Macedonia**

**Abstract:** Information and Communication Technology (ICT) affects to a great extent the output and productivity growth. Evidence suggests that investment growth in ICT has rapidly accelerated the TFP (total factor productivity) growth within the European Union. Such progress is particularly essential for the sectors which themselves produce new technology, but it is dispersing to other sectors, as well. Nevertheless, decrease in ICT investment does not necessarily decline the ICT contribution to output and productivity growth. These variations come out from the problems related to the particular phenomenon proper assessment, but predominantly from the companies' special requirements, as well as the necessary adjustments of labour employed. Hence, this paper aims at estimating the huge distinction in terms of ICT and TFB contributions to labour productivity growth among some of the European member states, as well as the factors which might stand behind the particular findings.

Key words: e-business, ICT, productivity, competitiveness

## 1. ICT as a factor of production

Different factors of production might affect GDP growth of the particular economy. The rise and fall of those which do not make clear the growth in production match to the general TFP achievement which is largely associated with the technical progress. Thus, capital and labor to output growth could be estimated by means of a flexible trans-log production function, such as:

$$\Delta Y = \bar{v}_k \Delta K + \bar{v}_l \Delta L + \Delta A \qquad \textbf{(1)}$$

where, $\bar{v}_k$ and $\bar{v}_l$ represent the input share in gross value added, while $\Delta A$ stands for the rise in output over the growth in weighted factor inputs or TFP growth (Jorgansen, Gollop and Fraumeni, 1987). Nevertheless, if capital input *k* is tried to be decomposed into three different types of ICT capital *c*, as well as the non-ICT capital *n* the equation above might be revised as follows:

$$\Delta Y = \sum_{i=o.m,s} \overline{v}_i^c \Delta k_i^c + \sum_{i=o,t,b} \overline{v}_i^n \Delta k_i^n + \overline{v}_i^l \Delta L + \Delta A \qquad \textbf{(2)}$$

where, variables $y$ and $k$ indicate the output ($Y$) per unit of labour input ($L$) and capital ($K$) per unit of labour input ($L$) respectively. Yet, TFP contribution to labor productivity could be additionally segregated into the possible gains from the ICT producing $A^c$ and other non-ICT industries $A^n$. The first ones stand for the technological change that follows the ICT production itself, while the second comprises the effects of ICT dispersal on other industries, as well as the other sources to TFP growth (Jorgansen and Stiroh, 2000). At the outset, ICT contribution to labor productivity growth ($\Delta Y^c$) might be re-estimated if one includes input-share weighted contributions of service flows from ICT assets $i$ within the total economy and the output-share weighted contributions of TFP in ICT producing industries $j$, or:

$$\Delta Y = \sum_i \overline{v}_i^c \Delta k_i^c + \sum_i \overline{u}_j^c \Delta A_j^c \qquad \textbf{(3)}$$

## 2. Assessing the ICT investment for several EU member states

Various methods are employed to replenish the breaks within the time series released on ICT investment in some EU countries (Schreyer, 2000) and (Davery, 2001, 2002). The most preferred, however, is the so called "commodity flow method" which traces commodities from imports or home production to the final procurement. For this purpose, input and output tables are usually united with data on office, communication and computer equipment.[1] Input and output tables (I/O), notwithstanding, correspond to domestic output and imports share preordained to investment. If one melds office, computer and communication equipment to investment, the following estimation might be obtained:

$$I_{i,t} = (Q_{i,t} - E_{i,t}^d)\left(\frac{I(Q)_{i,t}^{I/O}}{(Q - E^d)_{i,t}^{I/O}}\right) + (M_{i,t} - E_{i,t}^r)\left(\frac{I(M)_{i,t}^{I/O}}{(M - E^r)_{i,t}^{I/O}}\right) \qquad \textbf{(4)}$$

where, $I_{i,t}$ stands for investment in point $i$, within the year $t$, $Q_{i,t}$ represents the domestic output, $E_{i,t}^d$ symbolizes the exports from domestic production, $(Q - E^d)_{i,t}^{I/O}$ signify the home production for domestic use as from I/O tables, $M_{i,t}$ are the imports in year $t$, $E_{i,t}^r$ stands for the re-exports in $t$, $(M - E^r)_{i,t}^{I/O}$ corresponds to the imports excluding reexports as from I/O tables, while $I(M)_{i,t}^{I/O}$ denotes the investment originating from imports as from I/O tables.

Recent findings confirm that the three fundamental ICT categories (office and computer equipment, communication equipment and software), as well as the additional (non-ICT equipment, transport equipment and non-residential structure) accounted for 17.1% of EU Gross Fixed Capital Formation (GFCF) in 2000 (Figure 1).

---

[1] This method is not to be applied on software.

Nonetheless, the ICT share was 28.2% of GFCF, with the foremost part of distribution on computer and office machinery, particularly within Germany, Spain, Netherlands and United Kingdom, while Italy, Austria, Denmark and Sweden accounted for large shares in the software industry.

**Figure 1: Gross Fixed Capital Formation by category as-% share of total Non-Residential GFCF and Total equipment**

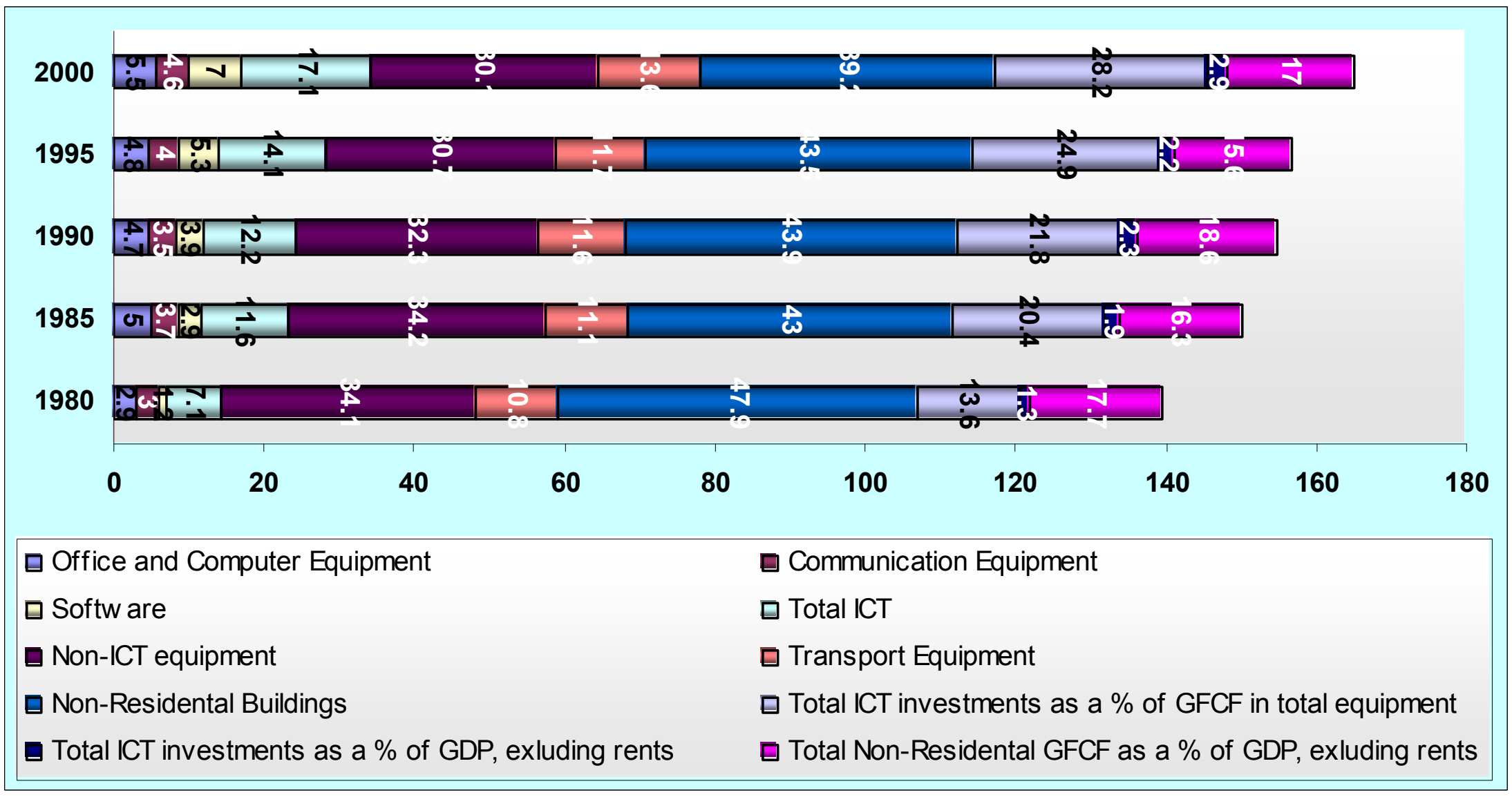

*Source:* Van Ark, Inklaar and McGuckin (2002, 2003a)

Noteworthy is to mention that growth rates of the real investment in communication and computer equipment has been even more rapid if one considers the deflators which reveal the EU price changes. In addition, ICT real investment growth was the most high-speeded in Ireland, which has started from relatively low level in the nineties, followed by the Nordic countries and Netherlands.

## 3. ICT conduciveness to output and labor productivity growth

Although the ICT contribution to annual labor productivity growth within several EU member states has been relatively high, the labor productivity growth itself has essentially set back. At the same time, TFP conduciveness to labor productivity growth has also slowed down starting from 1995.

### 3.1. ICT and labor productivity growth

Within the mid nineties aggregate labor productivity growth has held up in some European countries, not only in relative, but also in absolute terms (Figure 2). Evidence suggests that ICT capital contribution to labor productivity has been increased during the period 1995-2000 in comparison with the one by mid nineties. To some extent, the particular improvement has been a result of the enlarged conduciveness of computer and office machinery. Yet, large variations occur in absolute ICT contribution to labor productivity growth, in addition to its distribution among different ICT types of asset for a single country (McGuckin and Van Ark, 2002). Namely, United Kingdom, Ireland and Netherlands approach high contribution levels predominantly due to the elevated office and computer machinery, despite all the other countries which prove lower achievements in the particular asset type.

**Figure 2: Contribution of Total ICT Capital and TFP to Annual Average Labor Productivity Growth, 1980-2000 in EU countries**

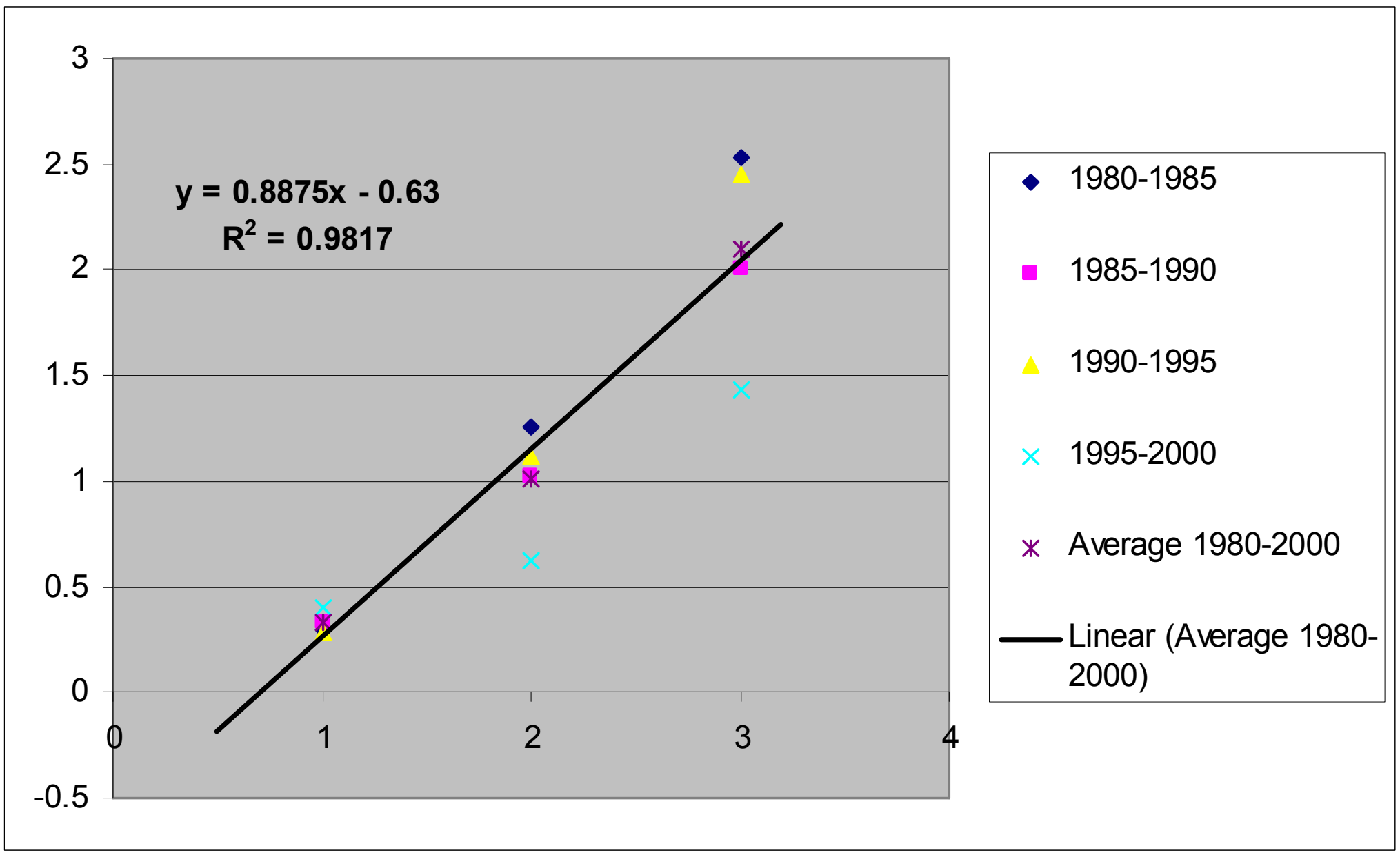

*Source:* Van Ark, Inklaar and McGuckin (2002, 2003a)

### 3.2. TFP and labor productivity growth

As mentioned above, ICT has a great contribution to output and labor productivity within some European economies. Besides the differentials in capital conduciveness, the major impact on labor productivity growth, however, has the total factor productivity growth. In consequence, labor productivity growth has been increased for the most part in Finland and Ireland since those are considered to be the major ICT producers, whereas United Kingdom and Netherlands suffered less augmentation although both are believed to be huge ICT investors. The equations proposed above indicate that TFP growth is dependant upon the differences in ICT producing industries which stand for the changes in technology, as well as the non-ICT industries which comprise the effects of ICT transmission to the other industries. The contributions of the both categories to aggregate TFP growth might be estimated by using the Domar final output weights. Thus, Evsey Domar (1961) has proved that aggregate TFP may possibly be rewritten as a weighted average of the particular industry productivity growth, whereupon the industry gross output – GDP ratio will be considered as the required weight, given as follows:

$$A_{GDP} = \sum_i \overline{w}_i A_i \qquad \textbf{(5)}$$

with $\overline{w}_i = \frac{1}{2}\left(\frac{GVO_{i,t}}{GDP_t} + \frac{GVO_{i,t-1}}{GDP_{t-1}}\right)$ **(6)**

where GVO stands for the value of gross final output of the industry i, $A_i$ represents the productivity growth of the industry i, while $A_{GDP}$ corresponds to the aggregate total factor productivity growth. The above estimations indicate that different concepts could be implemented at the industry and aggregate level (final output and value added, respectively). Additionally, the aggregate level comprises merely the primary inputs, while both primary and intermediary inputs are taken into the industry functions of production.

Evidence suggests that contributions of ICT-producing industries (office and computer machinery, communication equipment and semiconductors) to TFP growth have increased in about 40% within several EU member states (Figure 3). In other words, ICT manufacturers account for one third of TFP growth for the period of 1995-2000. Office and computer machinery, nevertheless, are to be the major contributors to TFP growth in United Kingdom, while Ireland additionally includes the semiconductors, as well. Recent findings indicate that TFP growth has gained from the communication equipment industry in Sweden and Finland, but with no predominance as sometimes proposed. Put forward differently, non-ICT producing industries account for the large share of TFP growth within the most of the EU countries, such as: Austria, Finland, Sweden and Ireland [2].

**Figure 3: Contributions of ICT-producing industries to TFP**

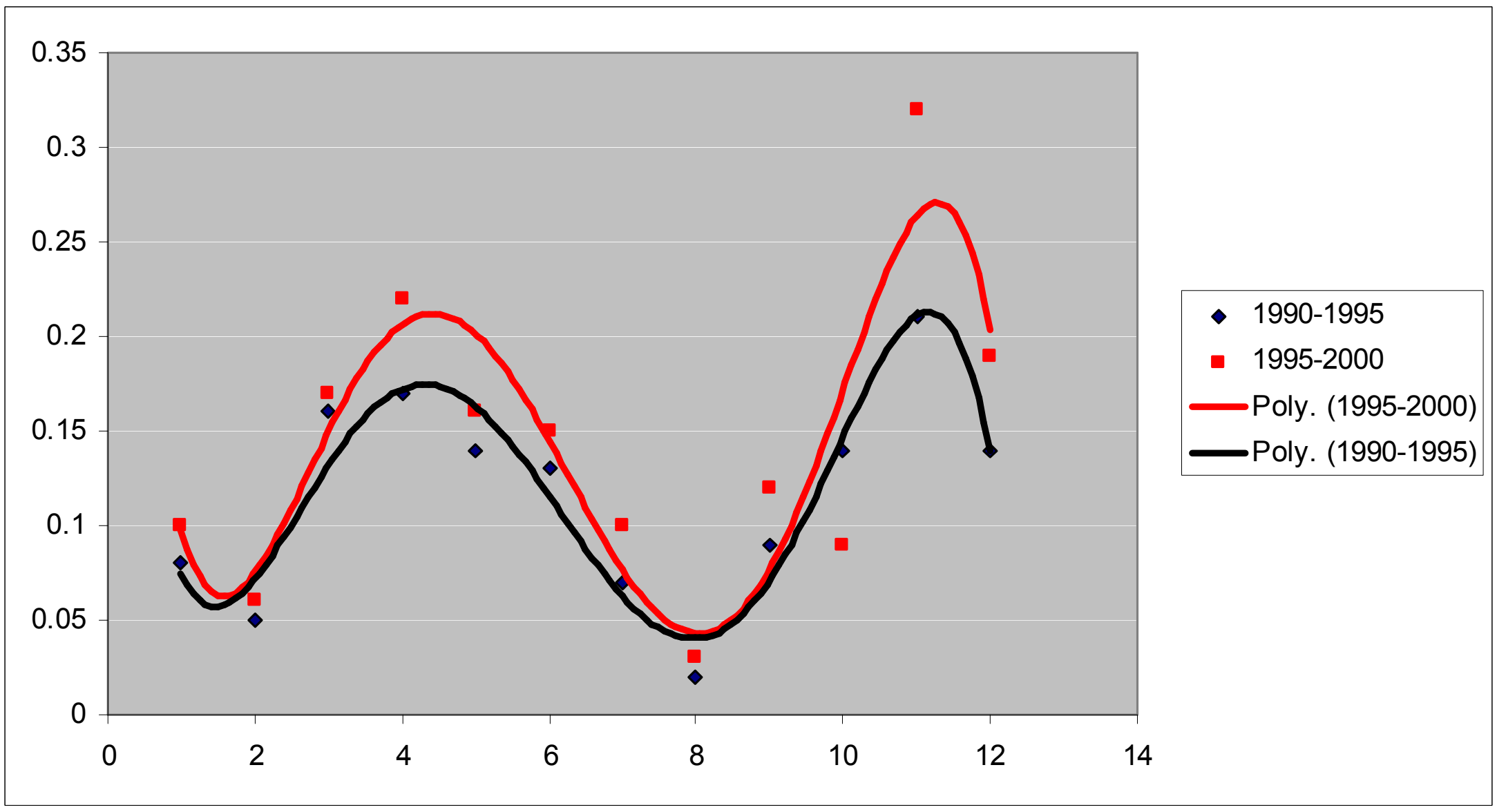

*Source:* Van Ark, Inklaar and McGuckin (2002, 2003a)

## 4. Conclusions

Within the early nineties European economies started with lower levels of ICT contributions to output and productivity growth, particularly because of the lesser production share in ICT. Many variations, however, emerge among different EU member states related to possible growth of ICT capital and the respective contributions to labor and TFP productivity growth. Thus, Netherlands, Ireland and

[2] The estimates for Ireland need some prerequisites, since the production shares are noticeably high. Thus, the ICT contribution to aggregate TFP growth might be computed by weighting the TFP growth rates in each industry with the particular Domar weight.

United Kingdom are being characterized by a large ICT conduciveness to productivity growth, while Spain and Portugal are likely to be at lower positions. On the other hand, Nordic countries are distinguished by the relative share of software as the main contributor to any considerable diffusion, but with no ample hastening of the productivity growth. Recent findings suggest that many European economies suffered productivity setback within the non-ICT service industry, despite those intensive ICT-using sectors. Nevertheless, ICT itself is not the only factor that has affected the particular productivity slowdown, but also the low levels of required skills, inflexible markets, the drop in capital/labor ratio etc. Many scholars (McGuckin, van Ark, 2001) argue that number of additional restraints hinder the ICT investment within the European Union, such as: regulations and structural impediments, restrictive labor rules and procedures, barriers to entry etc.